\documentclass[preprint,aps]{revtex4}
\usepackage{graphicx}
\newcommand{\be}{\begin{equation}}
\newcommand{\ee}{\end{equation}}
\newcommand{\ba}{\begin{eqnarray}}
\newcommand{\ea}{\end{eqnarray}}

\newcommand{\f}{\frac}

\begin{document}
\title
{The Rayleigh-Lorentz Invariant and Optimal Adiabatic 
Qubit-Information Detection for Superconducting Qubit Resonators
 }
\author{Jeong Ryeol Choi\footnote{E-mail: choiardor@hanmail.net } \vspace{0.3cm}}

\affiliation{Department of Electrophysics, 
Kyonggi University, 
Yeongtong-gu, Suwon,
Gyeonggi-do 16227, Republic of Korea \vspace{0.7cm}}

\begin{abstract}
\indent
Dynamical properties of a resonator can be analyzed using the Rayleigh-Lorentz invariant which
is not an exact constant but varies more or less over time.
We investigate the time behavior of this invariant for a flux qubit resonator
in order for better understanding of qubit-information detection with the resonator.
Flux qubit resonators can be utilized in 
implementing diverse next generation nano-optic and 
nano-electronic devices such as quantum computing systems.
Through the analyses of the temporal evolution of the invariant, 
we derive 
a condition for optimal adiabatic qubit-information detection
with the resonator.
This condition is 
helpful for controlling the dynamics of qubit resonators over long periods of
time.
It is necessary to 
consider it 
when designing a nano-resonator used for
quantum nondemolition readouts of qubit states, crucial in quantum computation.
\\

\end{abstract}

\maketitle

{\ \ \ } \\
{\bf 1.
Introduction 
\vspace{0.2cm}} \\
Flux qubit resonators \cite{df3,fqr1,fqr2} can potentially be implemented to diverse amplification schemes
for measuring weak information signals in quantum systems such as quantum computers.
Adiabatic readouts of qubit states in quantum computation using controllable 
resonators are required in optimized computing models \cite{p0,n1,p4},
while adiabatic quantum computation with a robust
quantum algorithm can be achieved on the basis of the adiabatic theorem \cite{dah,swg4}.
Due to this, adiabatic evolution of a resonator \cite{as0,n1} incorporated
with 
nondemolition qubit information detection 
has attracted considerable interest in quantum-information science.
It is possible to investigate dynamical properties of a flux qubit resonator from
the analysis of the variation of the associated 
Rayleigh-Lorentz adiabatic invariant \cite{lra}.

It is well known in quantum mechanics that, if the adiabatic hypothesis related
to the Rayleigh-Lorentz adiabatic invariants
holds, the initial eigenstate in the discrete spectrum of the Hamiltonian remains the same over time.
In order to process quantum information using computing algorithms and to read out
qubit-state signals, a high-fidelity resonator is indispensable.
As an
implementation of devices for such purposes,
a flux qubit is an artificial two-level system that is a basic unit that stores quantum information.
Typically, flux qubits are fabricated by superconducting circuits using nanotechnology facilities.
The entanglement between a flux qubit and a SQUID (superconducting quantum interference device)
is usually used as 
a protocol for measuring the quantum states \cite{flx}.

Adiabatic invariants
that are nearly conserved quantities
when the system parameters change slowly have been one of the core research subjects concerning time-varying mechanical systems.
After Burgers' pioneering work \cite{jmb} in 
the adiabatic hypothesis and its applications, 
adiabatic invariants for both nonconservative and nonlinear systems have been extensively
investigated \cite{awc,lls,rgo}. 
The reason why adiabatic invariants have become a topic of interest is that
we can deduce various dynamical properties of
a system from such conserved quantities, leading to deepening the understanding of the system.
Indeed, adiabatic invariants are useful for characterizing quantal and photonic properties for
adiabatically evolving nanosystems \cite{aai1,aai2,aai3,aai4,aai5}.

Such Rayleigh-Lorentz invariants are not exact constants, but approximate constants
under the assumption that the variations of parameters are sufficiently slow.
Namely, 
the Rayleigh-Lorentz invariants vary more or less with time. 
The study of such variation 
for specific systems may
allow us to gain insight in understanding the underlying
mechanism associated with the invariants \cite{lls}.
The mechanics of such adiabatic invariance can be applied to analyzing dynamical properties of
superconducting flux qubits in adiabatic quantum computation \cite{hda,swg4}.


In this work, we investigate the characteristics of the Rayleigh-Lorentz invariant of the
nano-resonator system and
find requirements for optimal qubit signal detection by utilizing such characteristics of the invariant.
To attain fault-tolerant quantum computation, computational states should not be
disturbed when we detect qubit information.
Hence, wave function of a qubit state should be precluded from undergoing a decoherence-induced collapse during its measurement.
In this regard, ideal qubit readout with 
high-fidelity is possible from a protective measurement
based on the preservation of adiabatic evolution 
of qubit eigenstates.
Protective measurement minimizes disturbance to the system and can possibly be used {\it in situ} as
a standard quantum measurement with reliable precision.
Based on our consequence for the condition for 
optimal adiabatic qubit-information detection,
we will address quantum nondemolition measurement which is critical
in order to extract scalable qubit information in 
quantum computation.
\\
\\
{\bf 2. 
Description of the Superconducting Qubit Resonator \vspace{0.2cm}} \\
While the picture of parametric resonance in a 
cavity is rich and rather complicated,
the designing of 
superconducting resonators is flexible thank to 
diverse available methods for parametric pumping \cite{df3}.
Hence, the characteristics of flux qubit resonators and
their mathematical representation are more or less different
depending on adopted models and fabrication methodologies.
Moreover, for a specific qubit resonator,
the degree of approximation for its nonlinear terms in the circuit also affects 
the explicit form of the equation of motion for the time behavior of a flux.
Throughout this work, we consider a kind of flux qubit resonator that were recently proposed 
and analyzed by Krantz {\it et al}. for convenience \cite{pkr,
kms,kra1}.
Krantz {\it et al}. adopted quarter wavelength superconducting resonators that include a coplanar waveguide
transmission line as a practical tool for reading out qubit states.
The response of this system near the resonance frequency can be modeled in terms of a parallel
RLC resonating circuit.
The phase difference in the SQUID is represented as $\varphi = 2\pi \phi /\phi_0$
where $\phi$ is the magnetic
flux in the superconducting loop while $\phi_0$ is the magnetic flux quantum which is given 
by $\phi_0 = \pi \hbar/e$.
Because 
the flux is quantized, the allowed quantities of $\varphi$ are discrete. 
When we describe complicated electronic circuits including Josephson junctions,
we can choose either charge $q$ or flux $\phi$ (or $\varphi$) as
coordinate. If we choose $q$ as coordinate, $\phi$ can be managed as the conjugate momentum,
while $q$ is regarded as momentum in the case where 
$\phi$ has been chosen to be coordinate.
Because the flux in Josephson junctions of the SQUID exhibits nonlinear characteristics, it is favorable
in this case to choose $\phi$ as coordinate \cite{rus}.
The resonator acts as a parametric oscillator that can be tunable by adjusting
the overall inductance (or capacitance).
For some technical reasons, the modulation of frequency by a nonlinear flux-tunable inductance is preferable to
tuning the capacitance \cite{pkr}.
The resonator can be operated by a radially oscillating small ac-flux
added to a static dc-flux. In many cases, the operated angular frequency (pumping frequency) of the ac-flux is nearly
twice the resonant angular frequency, $\omega_p \approx 2 \omega_r$.

For the tunable quarter wavelength flux qubit resonators considered both nonlinearity
and damping, an extended Duffing equation is given by \cite{kra2,kra1,df3}
\be
\f{d^2 \phi}{d t^2}+(\omega_r/Q) \f{d \phi}{d t} + \omega^2(t)\phi
-\alpha \Lambda(t)\phi^3= \xi(t) .
\label{2}
\ee
Here, 
$Q$ is the quality factor,
$\alpha$ is the Duffing nonlinear term,
$\xi(t)$ is a 
noise or external signal, $\omega(t)$ is a time-dependent angular
frequency, and $\Lambda(t)$ is a time function that is represented as
\be
\Lambda(t) =  1- 3\lambda Q \epsilon \cos (\omega_p t)/(2\omega_r \omega_p), \label{2dn}
\ee
where $\lambda$ is a correction to the Duffing nonlinearity caused 
by a modulation of $\alpha$
through the pump, and $\epsilon$ is the strength of pumping.
In Eq. (\ref{2}), higher order terms have been taken into account in order to meet experimental
results that exhibit pump-induced frequency shift.
Nonlinearity in the system has been
induced by connecting the SQUID to the cavity, whereas the damping takes place by
attaching the cavity to a transmission line.

In the representation of the flux equation, Eq. (\ref{2}), we have followed the convention of notations
as that in Ref. \cite{kra2}, which are expressed in terms of $\phi$, but if we rescale the Duffing nonlinear
term as $\alpha \rightarrow (2\pi/\phi_0)^2\alpha$, Eq. (\ref{2}) reduces to that in
Refs. \cite{kra1} and \cite{df3}.
The time behavior of the  Duffing
coefficient $\alpha \Lambda(t)$ characterized by Eq. (\ref{2dn}) is identical to that given in
Ref. \cite{df3} under an appropriate assumption, which is $|f(t)| \ll 1$ where $f(t)$ is a controlling field
defined in that reference. 
The most up-to-date technique for reading qubit data with high fidelity is using the nonlinear properties of
the nano-resonator coupled to the qubit system \cite{pkr}. 

The resonator system described by Eq. (\ref{2}) can be applied to various next-generation nanotechnologies
for quantum information processing.
For convenience, we consider a particular case that $\omega(t)$ is given in the form \cite{kra1}
\be
\omega(t) = \left[\omega_r^2+\epsilon \cos (\omega_p t)-
\f{\beta \epsilon^2Q}{2\omega_r \omega_p}(1-\cos(2\omega_p t))\right]^{1/2},
\label{3}
\ee
where $\beta$ is a dimensionless parameter.
If $\beta \rightarrow 0$, the frequency given in Eq. (\ref{3}) reduces to that of Eq. (1) in Ref. \cite{jkp}
and/or Eq. (1) in Ref. \cite{jkp1},
leading the system being 
similar to those treated in the same references.
On the other hand, the last term in the bracket of Eq. (\ref{3}) is the one that appeared in Eq. (1)
of Ref. \cite{mem}.
To know how to determine various parameters in the system, 
refer to Ref. \cite{pkr}.
For other models 
where the equation for the flux is different from Eq. (\ref{2}),
refer to Refs. \cite{df1} 
and
\cite{df6
}.


Now we consider the case of a weak Duffing nonlinear term, that can be established
by putting $\alpha \simeq 0$, as a solvable case.
Upon this situation, the Hamiltonian describing Eq. (\ref{2}) is a quadratic form and the corresponding
energy can be written as 
(see 
Appendix A) 
\be
E(t)=  \exp({-\omega_r t/Q})\f{\omega(t)}{\omega(0)}\left(E(0)+\f{C\xi^2(0)}{2\omega^2(0)}\right)-
\f{C\xi^2(t)}{2\omega^2(t)}, \label{4}
\ee
where $C$ is the capacitance of the resonator.
\\
\\
{\bf 3. 
Rayleigh-Lorentz Invariant \vspace{0.2cm}} \\
Rayleigh \cite{lra} discovered, for a specific time-varying system, that the 
quantity $I(t)$, which is defined as
\be
I(t) = {E(t)}/{\omega(t)}, 
\label{1}
\ee
almost does not vary
over time, provided that the variations of system parameters are sufficiently slow.
Subsequently, Lorentz 
rediscovered this consequence in the semiclassical regime and 
pronounced his discovery at the famous first Solvay Conference 
(for detailed reviews 
of this, see Refs. \cite{lls} and \cite{rgo}).

\begin{figure}
\centering
\includegraphics[keepaspectratio=true]{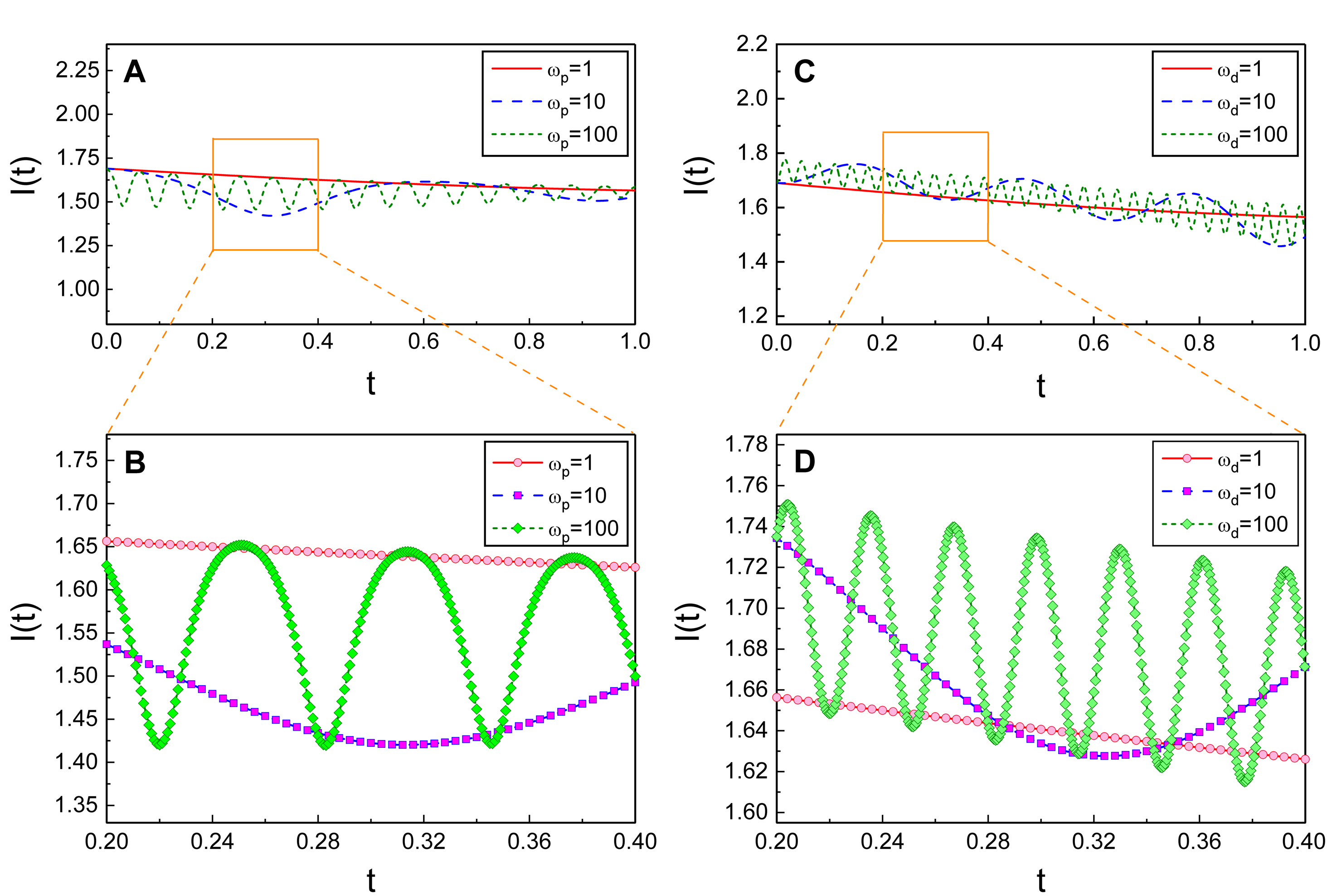}
\caption{\label{Fig1}
{The effects of the increase of the frequencies $\omega_p$ and $\omega_d$
on temporal evolution of the Rayleigh-Lorentz invariant,} where a sinusoidal 
noise $\xi(t) = \xi_0\cos(\omega_d t+\theta)$ has been taken.
Here, $\xi_0$ is the amplitude of the external force and $\omega_d$ is the driving frequency.
(A) is for several different values of $\omega_p$, whereas (C) for $\omega_d$.
We have used $\omega_d=1$ for (A) and $\omega_p=1$ for (C).
Other quantities that we have used are $\omega_r=0.5$, $\epsilon=0.1$, $\beta=1$, $Q=5$, $\xi_0=0.2$,
$E(0)=1$, $C=1$, and $\theta=0$. (B) and (D) are enlarged plots
between $t=0.2$ and $t=0.4$ for (A) and (C), respectively.}
\end{figure}

If we insert Eq. (\ref{4}) in Eq. (\ref{1}), we obtain the Rayleigh-Lorentz invariant of the system
such that
\be
I(t) = \f{\exp({-\omega_r t/Q})}{\omega(0)}\left(E(0)+\f{C\xi^2(0)}{2\omega^2(0)}\right)-
\f{C\xi^2(t)}{2\omega^3(t)}, \label{5}
\ee
where $\omega(t)$ is given by Eq. (\ref{3}).
Apparently, 
the ratio of energy to the angular frequency is an adiabatic invariant
that is useful for studying dynamical properties of the system \cite{as0}.
The adiabatic invariant given in Eq. (\ref{5}) is an approximate constant under the condition that
the variations of parameters of the dynamical system are sufficiently slow.
\begin{figure}
\centering
\includegraphics[keepaspectratio=true]{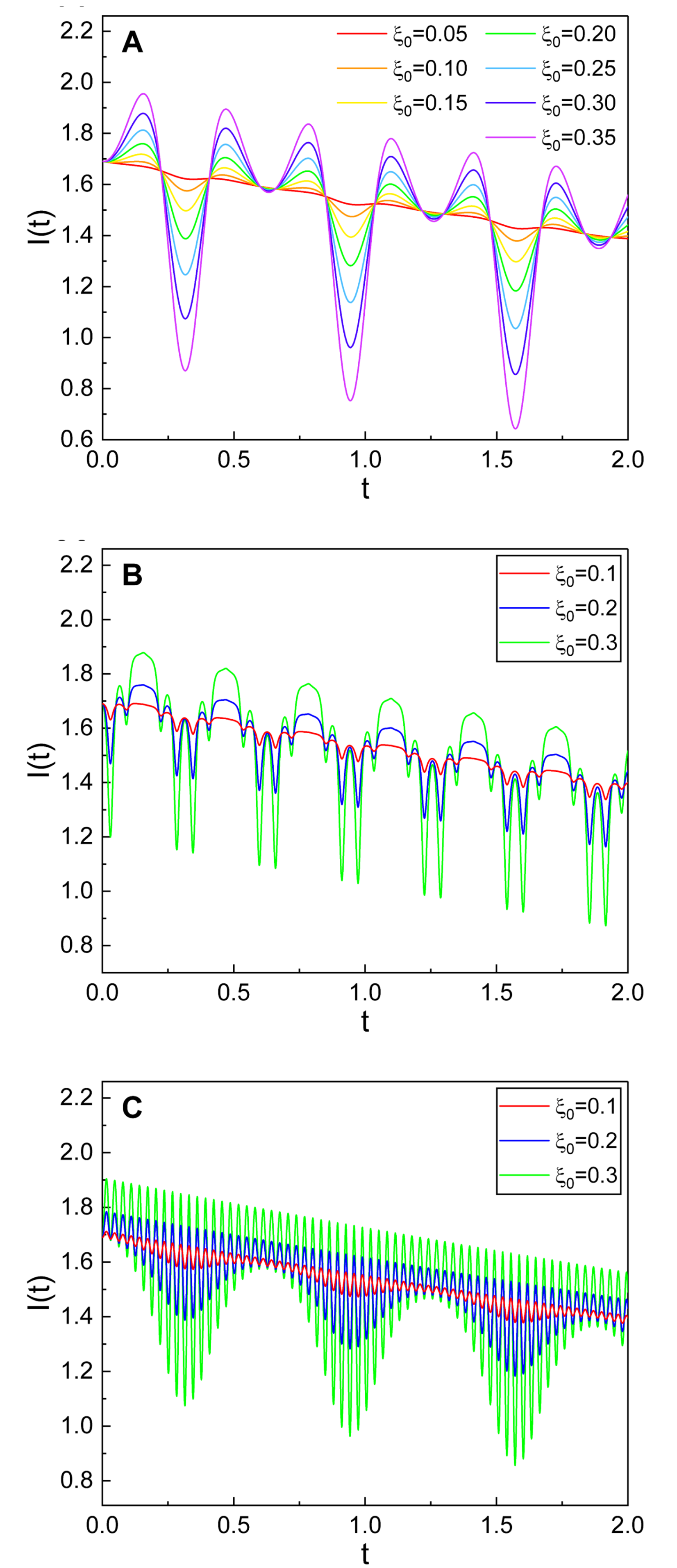}
\caption{\label{Fig2}
Emergence of large-scale oscillations of the Rayleigh-Lorentz invariant via the joint
effects of $\omega_p$ and $\omega_d$, which is shown for several different values of $\xi_0$.
The same formula of $\xi(t)$ as that in Fig. 1 is taken.
The values of ($\omega_p$, $\omega_d$) that we have used are
(10, 10) for (A), (100, 10) for (B), and (10, 100) for (C).
Other values are common and given by $\omega_r=0.5$, $\epsilon=0.1$, $\beta=1$,
$Q=5$, $E(0)=1$, $C=1$, and $\theta=0$.
}
\end{figure}
\begin{figure}
\centering
\includegraphics[keepaspectratio=true]{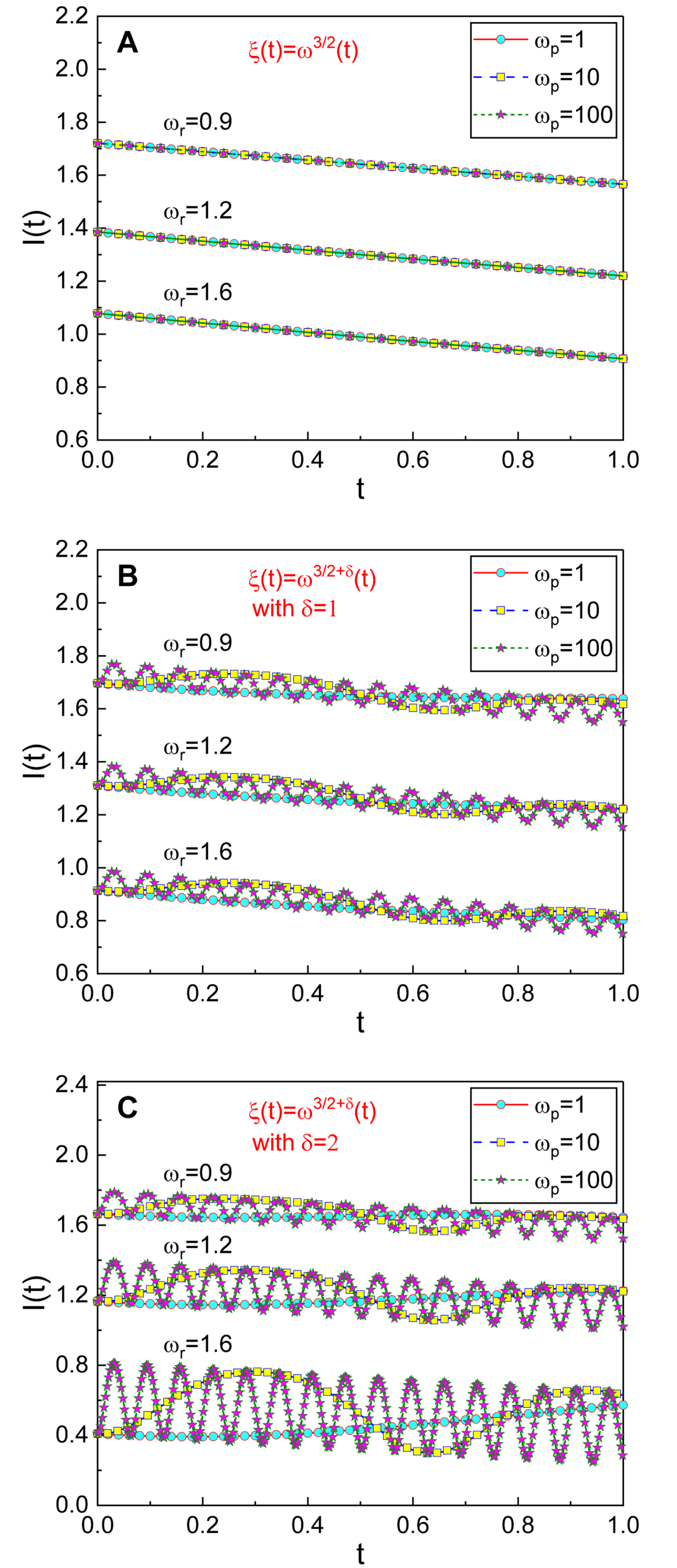}
\caption{\label{Fig3}
{Graphical demonstration 
for optimal adiabatic condition}
from temporal evolution of $I(t)$.
Each panel is drawn for several different values of $\omega_p$ and $\omega_r$.
$\xi(t) = \xi_0\omega^{3/2+\delta}(t)$ has been taken
regarding Eq. (\ref{p5+1}), where $\delta$ is a deviation from the optimal condition.
Panel (A) is for $\delta = 0$, (B) for $\delta = 1$,
and (C) for $\delta = 2$.
The values of parameters that we have taken are $\epsilon=0.5$, $\beta=0.5$,
$Q=10$, $\xi_0=0.4$, $E(0)=2$, and $C=1$.
(A) shows that $I(t)$ does not vary locally when $\xi(t)$ follows the condition in Eq. (\ref{p5+1}).
If $\xi(t)$ deviates from that condition, the fluctuation of $I(t)$ emerges (B, C).}
\end{figure}

Let us analyze the time behavior of  $I(t)$ for several particular cases.
We have plotted its temporal evolution 
using 
Eq. (\ref{5}) in Fig. 1 for several different
choices of the value of parameters.
We have chosen $\xi(t)$ as a sinusoidal form in these analyses for the purposes
of simplicity. 
While $I(t)$ almost does not vary for the case of small values of $\omega_p$ or $\omega_d$,
it oscillates as $\omega_p$ and/or $\omega_d$ become large.
S$\acute{\rm a}$nchez-Soto and Zoido discovered similar oscillations of $I(t)$ for the systems of
linear or exponentially lengthening pendulums \cite{lls}.
Figure 2 shows large-scale oscillation of $I(t)$ through the joint effects of
$\omega_p$ and $\omega_d$.
This figure also exhibits the fact that the amplitudes of such oscillations become high as $\xi_0$
increases.
We can confirm from these analyses that, if the process of the change for
the parameters of the
system 
is
too fast, $I(t)$ would not remain constant.
\\
\\
{\bf 4. 
Optimal Adiabatic Condition \vspace{0.2cm}} \\
Let us deduce a useful adiabatic condition between system parameters from the formula of Eq. (\ref{5}).
The first term in Eq. (\ref{5}) exponentially decays out as time goes by.
Hence, it vanishes for a sufficient large $t$ and, as a consequence, we obtain
a 
useful parametric behavior from the remaining term, which yields at later time, as
\be
\xi(t) \propto \omega^{3/2}(t). \label{p5+1}
\ee
This is the main consequence of the present work.
When controlling dynamics of the resonator adiabatically \cite{as0,a1} over a long time, one should
consider this relation. By comparing Fig. 3(A) with Figs. 3(B,C) (or with other previous figures),
we see that the variation of the invariant
is negligible in the case where this condition has been met.
If we regard that one of the main problems for quantum computing is achieving high-fidelity quantum
nondemolition readouts of the qubit states without measurement-induced decoherence,
protective or nondemolition measurement of qubit states upon adiabatic approximation is
important \cite{qnm0,qnm}.
Quantum-limited metrology
in quantum computing systems requires reliable state detection with high efficiency.
Practical implementation of protective quantum measurements is based on the fact that,
while we can completely describe a system using the Schr\"{o}dinger wave,
the quantum state in adiabatic measurement does not change throughout the experiment.
\\
\\
{\bf 5. 
Conclusion 
\vspace{0.2cm}} \\
In light of the present research, the invariant plays a crucial role in investigating
the dynamics of information detection in the flux qubit resonator.
The invariant is approximately constant for small values of the pumping
and driving frequencies. We see that the first time-dependent term of the invariant given in Eq. (\ref{5}) decays
exponentially. Hence, the last 
term in the same equation
should become approximately constant over
sufficiently long 
periods.
In this way, 
a 
useful 
condition for adiabaticity has been found under which the invariant is
approximately constant over estimated long-time scales, as shown in Eq. (\ref{p5+1}).
This condition could help to control the dynamics of the flux qubit resonator over 
long periods of time.
Rigorous conservation of adiabatic invariants is requisite in order to keep the system
being adiabatic during the operation of the nano-resonator \cite{tsu,tsu2}.
Hence, 
Eq. (\ref{p5+1}) is important as a requirement 
for optimal qubit-information detection in
protective/nondemolition adiabatic measurements \cite{p0,
p4,
n1,as0,a1,
qnm0,qnm
}.

In order to prevent liable transfer of the 
eigenstate of the system Hamiltonian to other ones,
it is necessary to preserve adiabaticity of the eigenstate.
Then the collapse or entanglement of the system would not appear and the eigenstate
can be measured with a high precision.
Protective measurement is possible in such a way, which helps in detecting the actual physical state characterized
by the wave function for a quantum system.
The temporal response of qubit readout is well-characterized and qualitatively
understood from adiabatic measurement, while high fidelity readouts of a qubit state are of
central importance for achieving a successful realization of quantum computers.
\\
\\
{\bf 
Appendix A: Derivation of the energy expression 
\vspace{0.2cm}} \\
The method for deriving energy expression given in Eq. (\ref{4}) appears in Ref. \cite{kra2}.
We briefly review it
starting from the general representation of the energy of the system: 
\be
E(t) = e^{-2\omega_r t/Q} \f{q^2}{2C}  + \f{1}{2}C\left[ \omega^2(t) \phi^2
- 2\xi(t) \phi \right], \label{A1}
\ee
where $q$ is the charge stored in $C$.
Let us consider the classical action 
of the form, $J = \oint q d \phi$.
Then the integration using Eq. (\ref{A1}) 
gives 
\be
J(t) = \f{2\pi e^{\omega_r t/Q}}{\omega(t)}\bigg( E(t)+\f{C\xi^2(t)}{2\omega^2(t)} \bigg). \label{7}
\ee
Hence, $J$ has been represented in terms of
$E(t)$. Now from the relation $J(t) = J(0)$, 
we easily have the formula of the energy given in Eq. (\ref{4}). 
\\


\end{document}